\documentclass[a4paper,11pt]{article}
\pdfoutput=1
\usepackage{jheppub}
\usepackage{color}
\usepackage{array}
\newcolumntype{C}[1]{>{\centering\arraybackslash}p{#1}}
\usepackage{slashed,dsfont}
\usepackage{bm,wasysym}
\usepackage{longtable}
\usepackage{booktabs}
\usepackage{slashed}

\usepackage{tabularx}
\usepackage{rotating}
\usepackage{lscape} 
\usepackage{hyperref}
\usepackage[normalem]{ulem}
\usepackage{braket}
\newcommand{\lsim}{
\mathrel{\hbox{\rlap{\hbox{\lower4pt\hbox{$\sim$}}}\hbox{$<$}}}}

\newcommand{\gsim}{
\mathrel{\hbox{\rlap{\hbox{\lower4pt\hbox{$\sim$}}}\hbox{$>$}}}}

\renewcommand{\arraystretch}{2}

\allowdisplaybreaks[2]

\newcommand{\nn}{\nonumber}

\def\re{{\rm Re}}  \def\im{{\rm Im}}

\def\mathB#1{{\mathcal{B}_{{#1}}}}
\def\mB#1{{m_{\mathcal{B}_{{#1}}}}} 
\def\pB#1{{p_{\mathcal{B}_{{#1}}}}} 
\def\mmB#1{{m^2_{\mathcal{B}_{{#1}}}}} 

\definecolor{schrift}{RGB}{120,0,0}

\title{\boldmath\color{schrift}{CP violation with GeV-scale Majorana neutrino in $\Lambda_b\to(\Lambda_c^+, p^+)\pi^+\mu^-\mu^-$ decays}}

\author{Diganta Das, Jaydeb Das}
\affiliation{{\sf Department of Physics and Astrophysics, University of Delhi, Delhi 110007, India}}
\emailAdd{diganta99@gmail.com}
\emailAdd{jaydebphysics@gmail.com}

\abstract{We explore the possibility of CP violation in baryonic $\Lambda_b\to(\Lambda_c^+, p^+)\pi^+\mu^-\mu^-$ decays which are mediated by two Majorana sterile neutrino and are $|\Delta L|=2$ lepton number violating processes. Appreciable CP asymmetry can be obtained if there are two on-shell Majorana neutrinos that are quasi-degenerate in mass with the mass difference of the order of average decay widths. We find that given the present constraints on the heavy to light mixing element $|V_{\mu N}|$, the $\Lambda_b\to p^+\pi^+\mu^-\mu^-$ and $\Lambda_b\to \Lambda_c^+\pi^+\mu^-\mu^-$ decay rates are suppressed but could be within the experimental reach at the LHC. If searches of the modes are performed, then experimental limits on the rates can be translated to constraints on the Majorana neutrino mass $m_N$ and heavy to light mixing element squared $|V_{\mu N}|^2$. We show that the constraints on the $(m_N, |V_{\mu N}|^2)$ parameter space coming from the $|\Delta L| = 2$ baryonic decays are complementary to the bounds coming from other processes.} 


\begin{document}

\maketitle

\renewcommand{\arraystretch}{1.6}

%%%%%%%%%%%%%%%%%%%%%%%%%%%%%%%%%%%%%%%%%%%%%%%%%%%%%%%%%%%%%%%%%%%
%%%%%%%%%%%%%%%%%%%%%%%%%%%%%%%%%%%%%%%%%%%%%%%%%%%%%%%%%%%%%%%%%%%
\section{Introduction} 
The neutrino oscillation experiments confirm that at least two of the three active light neutrinos are massive \cite{Fukuda:1998mi,Wendell:2010md,Ambrosio:2003yz}. This opens up the possibility of CP violation in the leptonic interactions which can be searched in neutrino oscillation experiments \cite{Cabibbo:1977nk}. Leptonic CP violation can arise in the same manner as in the quark sector, namely complex phases in the leptonic mixing matrix. Whether the neutrinos are of the Dirac or Majorana type, CP violation is expected in both cases. But two additional sources of CP-violating phases can arise if the neutrinos are Majorana rather than if they are Dirac. Majorana character plays an important role as far the origin of the smallness of the active neutrino masses are concerned. If $N_R$ is a Standard Model right-handed gauge-singlet (and hence sterile) neutrino, then the Standard Model allows both Dirac mass term of the type $m_D(\overline{\nu}_L N_R + {\rm h.c})$, and a Majorana term of the type $m_N N_R N_R$. Then, via `see-saw' mechanism one can have small active neutrino mass $m_{\nu }\sim m_D^2/m_N$ if $m_D$ is at the electroweak scale or lower \cite{Mohapatra:1979ia,Schechter:1980gr,Schechter:1981cv,Minkowski:1977sc,Yanagida:1979as,Ramond:1979py,Levy:1980ws}. In the simplest version of the mechanism, the so-called type-I see-saw, the heavy electroweak singlet $N_R$ is of few TeV are introduced that give rise to the light eigenstates $m_\nu\lesssim 1$eV. However, low energy see-saw mechanism, where the sterile states $N_R$ are in the range of few hundreds of MeV to a few GeV, also have been proposed \cite{Buchmuller:1991ce,Asaka:2005an,delAguila:2007ap,He:2009ua,Kersten:2007vk,Ibarra:2010xw,Nemevsek:2012cd}. These so-called GeV-scale sterile neutrinos have several advantages: they could simultaneously explain the baryon asymmetry of the universe \cite{Asaka:2005an,Asaka:2005pn,Canetti:2014dka,Canetti:2012kh,Shuve:2014zua}, and can be experimentally searched both at the intensity and the energy frontier.

An important distinguishing feature between Dirac and Majorana sterile neutrino is that the later participates in $|\Delta L|=2$ lepton number violating (LNV) decays. For a light Majorana exchange, the neutrino-less double beta decay ($0\nu\beta\beta$) \cite{Pas:2015eia,Rodejohann:2011mu,DellOro:2016tmg,GomezCadenas:2011it} is one of the most sensitive probe of lepton number violation. But it was recently pointed out that with the exchange of heavy Majorana neutrino at the GeV scale, this rate can be enhanced \cite{Drewes:2016lqo, Asaka:2016zib}. Unfortunately, $0\nu\beta\beta$ process is yet to be experimentally verified and the best limit on the half-lives of different isotopes ($^{76}{\rm Ge}$, $^{136}{\rm Xe}$, $^{130}{\rm Te}$) come from several different experiments \cite{Aalseth:2017btx,Agostini:2018tnm,Alduino:2017ehq,Albert:2017owj,KamLAND-Zen:2016pfg}. Due to the lack of evidence of LNV decay so far, it is imperative to pursue complementary search strategies. This is further reinforced by the fact that observation of $0\nu\beta\beta$ only confirms lepton number violation in the first family of neutrinos, and to observe the same in other families, alternative processes must be investigated. Lepton number violating rare decays of mesons and baryons which are mediated by Majorana neutrino is important in this regard. For light or heavy Majorana neutrino exchange, decay rates are too suppressed to be accessed by current experiments. But, if the Majorana mass is within a few hundred MeV to a few GeV then the decay rates can be within the sensitivity reach of future experiments \cite{Helo:2010cw, Cvetic:2010rw}. Due to ongoing searches of LNV processes at flavor factories including the LHC and Belle-II, there have been theoretical interests in the LNV decays of hadrons \cite{Helo:2010cw, Cvetic:2010rw, Atre:2005eb, Dib:2000wm, Ali:2001gsa, Zhang:2010um, Yuan:2013yba, Godbole:2020doo, Cvetic:2020lyh, Shuve:2016muy, Chun:2019nwi, Mandal:2017tab, Abada:2017jjx, Abada:2019bac, Mejia-Guisao:2017nzx, Zhang:2021wjj, Cvetic:2019shl, Barbero:2013fc, Mandal:2016hpr, Zamora-Saa:2016qlk,Cvetic:2015ura,Cvetic:2015naa,Cvetic:2014nla,Cvetic:2013eza,Kim:2018uht,Kim:2019xqj,Milanes:2018aku,Mejia-Guisao:2017gqp,Milanes:2016rzr,Castro:2013jsn,Quintero:2011yh,Littenberg:1991rd,Barbero:2002wm}, $\tau$-lepton decays \cite{Castro:2012gi,Dib:2011hc,Yuan:2017xdp,Zamora-Saa:2016ito,Kim:2017pra}, and in different scattering processes \cite{Das:2017zjc,Das:2017rsu,Das:2017nvm,Cvetic:2019rms,Cvetic:2018elt,Fuks:2020zbm,Fuks:2020att,Cai:2017mow,Ruiz:2020cjx,Najafi:2020dkp}. The LHCb has searched for the process $B^-\to\pi^+\mu^-\mu^-$ \cite{Aaij:2014aba} and the NA48/2 has searched for $K^-\to\pi^+\mu^-\mu^-$ \cite{CERNNA48/2:2016tdo} and these experiments provide stringent constraints on the heavy to light mixing matrix elements. With large integrated luminosity coming from the Belle-II as well as upgrade of the LHCb, sensitivity to $|\Delta L|=2$ processes in mesons and baryons is expected to increase.

In this paper we study lepton number violating four-body $\mathcal{B}_1\to \mathcal{B}_2^\mp\pi^\mp\ell_1^\pm\ell_2^\pm$ decay, where $\mathcal{B}_1^0$ is $\Lambda_b$ and $\mathcal{B}_2^+$ is either a $\Lambda_c^+$ or $p^+$, and $\ell_1$ and $\ell_2$ can in general be of different flavors. Previously, in Ref.~\cite{Mejia-Guisao:2017nzx} these decays were considered in a model involving single on-shell Majorana exchange at the GeV scale. We are interested in a scenario where the decays are mediated by the exchange of two almost degenerate Majorana neutrinos of mass in the range between a few hundred MeV to a few GeV so that they can be on-shell. An interesting consequence of two Majorana exchange is the possibility of CP violation. We show that the CP violation can be appreciable if the two Majoranas are almost degenerate with the mass difference of the order of decay widths, $\Delta m_N\sim\Gamma_N$. There are well-motivated models where quasi degenerate Majorana neutrinos in the range of few hundreds of MeV to few GeV are predicted \cite{Dib:2014pga}. We calculate the branching ratios for $\Delta m_N\sim\Gamma_N$ and find that for the present experimental bound on $|V_{\mu N}|^2$, the $\Lambda_b\to (\Lambda_c^+,p^+)\pi^+\mu\mu$ rates might be within the reach of LHC in the future. Even if the modes are not immediately seen, experimental limits on the decay rates can be used to obtain constrain on the neutrino mass $m_N$ and the neutrino mixing matrix elements $|V_{\mu N}|^2$.

The paper is organized as follows. In section \ref{sec:decay} we work out the formalism for a generic $\mathcal{B}_1\to \mathcal{B}_2^\mp\pi^\mp\ell_1^\pm\ell_2^\pm$ decay mediated by on-shell Majorana neutrino. In section \ref{sec:num} we perform a numerical analysis of the CP asymmetry for $\Lambda_b\to (\Lambda_c,p)\pi\mu\mu$, and discuss the constraint on the $(m_N, |V_{\mu N}|^2)$ parameter space assuming experimental upper limits. We summarize our results in sec \ref{sec:summary}. Some details of our derivations are given in the appendixes. 

%%%%%%%%%%%%%%%%%%%%%%%%%%%%%%%%%%%%%%%%%%%%%%%%%%%%%%%%%%%%%%%%%%%%%%
%%%%%%%%%%%%%%%%%%%%%%%%%%%%%%%%%%%%%%%%%%%%%%%%%%%%%%%%%%%%%%%%%%%%%%
\section{$\mathcal{B}_1\to \mathcal{B}_2^\mp\pi^\mp\ell_1^\pm\ell_2^\pm$ formalism \label{sec:decay}}
We consider a model scenario where in addition to the components $\nu_{\ell L}$ of the left-handed $SU(2)_L$ doublets of the Standard Model, there are two right-handed singlet sterile neutrinos denoted by $N_1, N_2$. The flavor eigenstates $\nu_{\ell L}$ can be written in terms of the mass eigenstates as
\begin{equation}
\nu_{\ell L} = \sum_{i=1}^3 U_{\ell i}\nu_{iL} + V_{\ell N_1} N_1 + V_{\ell N_2} N_2\, ,
\end{equation}
where $\nu_{iL}$ are the light mass eigenstates. We assume that the heavy to light mixing elements $V_{\ell N_1}$ and $V_{\ell N_2}$ are free parameters and can be constrained by experiments. They in general can be complex 
\begin{equation}\label{eq:phase}
V_{\ell N_j} = |V_{\ell N_j}| e^{i\phi_{\ell j}}\, ,\quad (j=1,2) \, ,
\end{equation}
where $\phi_{\ell j}$ is a CP-odd phase. According to our convention, $V_{\ell N}$ is the mixing element between negatively charged lepton $\ell$ and Majorana neutrino $N$.

\begin{figure}[h!]
	\begin{center}
		\includegraphics[scale=0.45]{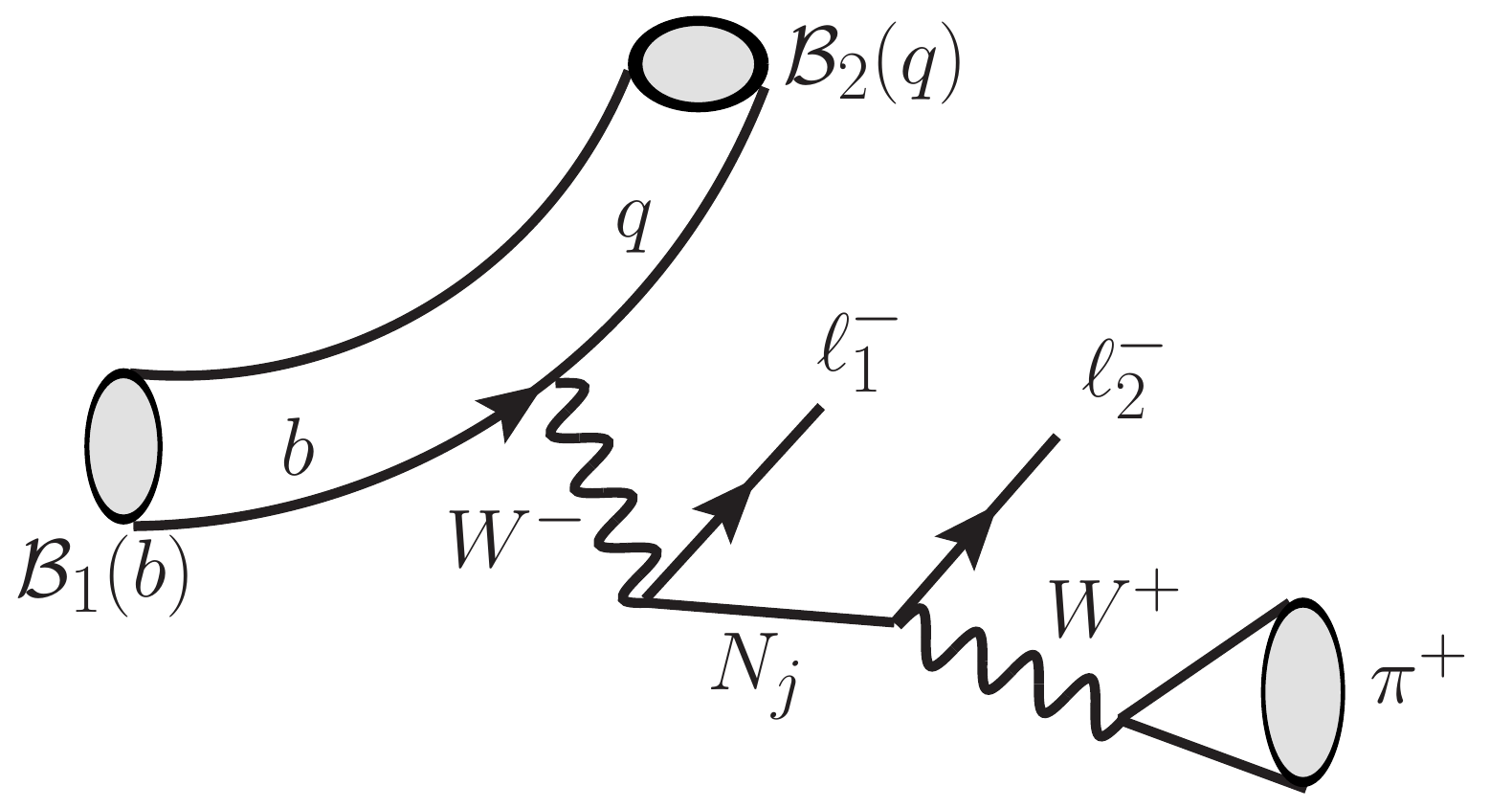}
		\includegraphics[scale=0.45]{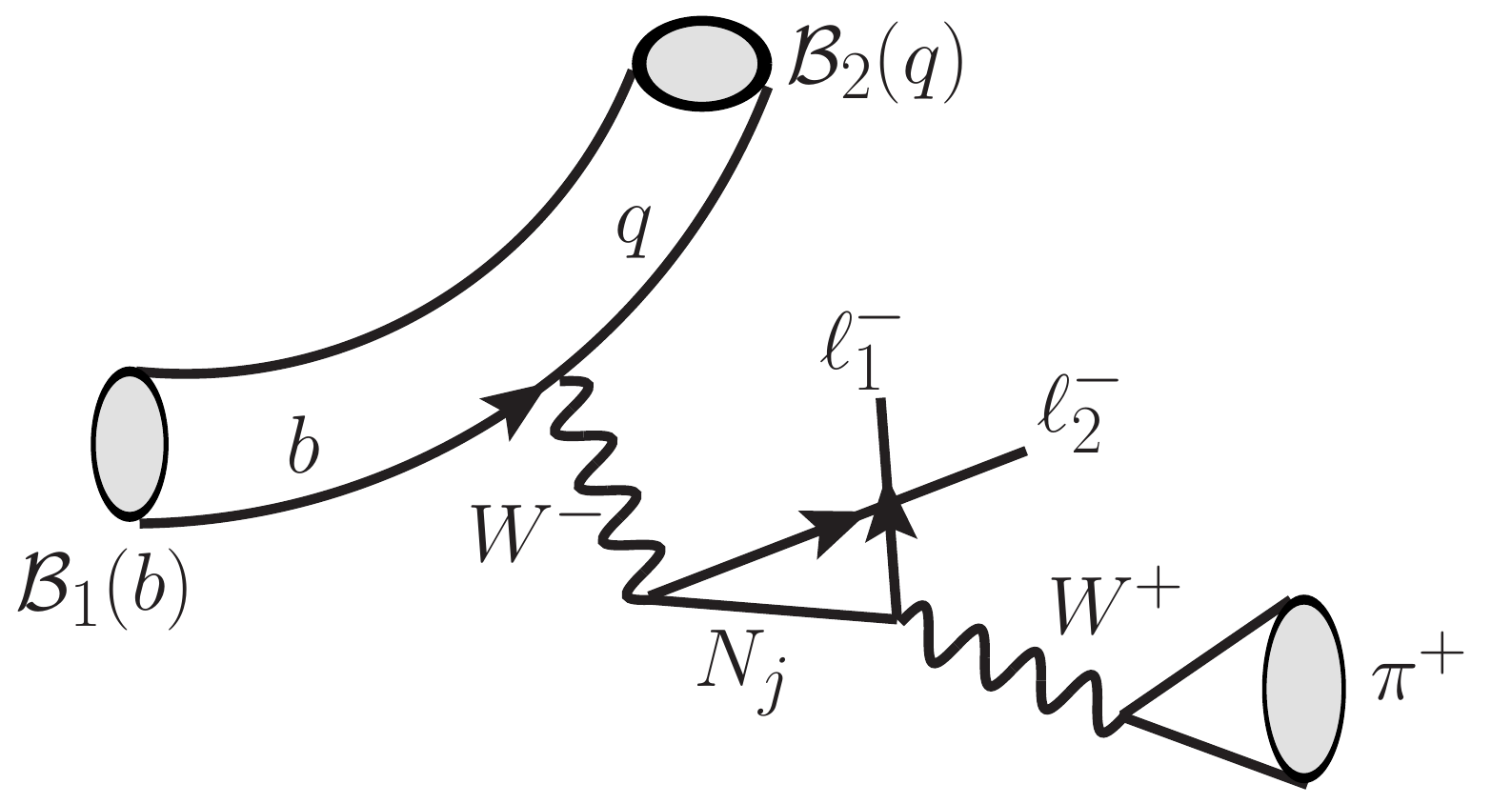}
		\caption{The direct ($D$) and cross($C$) channel Feynman diagrams for $\mathcal{B}_1\to \mathcal{B}_2^\mp\pi^\mp\ell_1^\pm\ell_2^\pm$ decay.  \label{fig:Feyn}}
	\end{center}
\end{figure}

We are interested to calculate the decay widths of $\mathcal{B}_1(\pB1)\to \mathcal{B}_2(\pB2)\pi^+(p_\pi)\ell_1^-(p_1)\ell_2^-(p_2)$ and its CP conjugate mode $\bar{\mathcal{B}}_1(\pB1)\to \bar{\mathcal{B}}_2(\pB2)\pi^-(p_\pi)\ell_1^+(p_1)\ell_2^+(p_2)$ in this model. The decays can be viewed as a two step processes: first, the $\mathcal{B}_1$ decays via a charged current interaction $\mathcal{B}_1\to \mathcal{B}_2^\mp N_j\ell_1^\pm$, followed by the decay of the heavy neutrino $N_j\to\ell_2^\pm \pi^\mp$. For these processes there are two dominant ``s-channel'' topologies, the direct channel ($D$) and the crossed channel ($C$), as shown in figure \ref{fig:Feyn}. The ``t-channel'' topologies are expected to be suppressed and are neglected. Appreciable decay rates can be obtained if the neutrinos have kinematically allowed mass
\begin{equation}
m_\pi + \ell_1 < m_{N_j} < (\mB{1}-\mB{2}-\ell_2)\, ,\quad \text{or/and}\quad m_\pi + \ell_2 < m_{N_j} < (\mB{1}-\mB{2}-\ell_1)\,.
\end{equation}
We denote the momentum of the heavy neutrino in the $D$ channel by $p_N = \pB1-\pB2-p_1$, and for the $C$ channel by $p^\prime_N = \pB1-\pB2-p_2$. Defining $\Gamma_{\mathcal{B}_1} \equiv \Gamma(\mathcal{B}_1\to \mathcal{B}_2\pi^+\ell_1^-\ell_2^- ) $ and $\Gamma_{\overline{\mathcal{B}}_1} \equiv \Gamma(\bar{\mathcal{B}}_1\to \bar{\mathcal{B}}_2\pi^-\ell_1^+\ell_2^+)$ the decay widths can be written as
\begin{equation}\label{eq:decay}
\Gamma_{\mathcal{B}_1(\overline{\mathcal{B}}_1)} = (2-\delta_{\ell_1\ell_2}) \frac{1}{2!} \frac{1}{2\mB1}\int d_4^{\rm PS} \bigg|\overline{\mathcal{M}}_{\rm tot}^{+(-)} \bigg|^2\, .
\end{equation}
The symmetry factor $1/2!$ comes because the two charged leptons can be the same. The $|\overline{\mathcal{M}}_{\rm tot}^{+}|^2$ ($|\overline{\mathcal{M}}_{\rm tot}^{-}|^2$) is the total matrix element mod-squared of $\mathcal{B}_1\to \mathcal{B}_2\pi^+\ell_1^-\ell_2^-$ $(\bar{\mathcal{B}}_1\to \bar{\mathcal{B}}_2\pi^-\ell_1^+\ell_2^+)$ after averaging over the initial spin and summing over the final spins
\begin{align}
|\overline{\mathcal{M}}_{\rm tot}^\pm |^2&=  \frac{1}{2}	\sum_{spins}\bigg|\mathcal{M}_{\rm tot}^\pm \bigg|^2\, \nn\\ 
&= \frac{1}{2}	\sum_{spins}\bigg|\sum_{j=1}^2\big(\mathcal{M}_{D_j}^\pm + \mathcal{M}_{C_j}^\pm\big) \bigg|^2\, ,\nn\\
& = \frac{1}{2}	\sum_{spins}\bigg[\sum_{i,j=1}^2 \mathcal{M}_{D_i}^\pm (\mathcal{M}_{D_j}^{\pm})^\ast + \sum_{i,j=1}^2 \mathcal{M}_{C_i}^\pm (\mathcal{M}_{C_j}^{\pm})^\ast + \sum_{i,j=1}^2 \mathcal{M}_{D_i}^\pm (\mathcal{M}_{C_j}^{\pm})^\ast + \sum_{i,j=1}^2 \mathcal{M}_{C_i}^\pm (\mathcal{M}_{D_j}^{\pm})^\ast \bigg]\, ,\nn\\ \label{eq:Mtot}
& =\mathcal{N}\bigg[ \sum_{i,j=1}^2 v_i^\pm (v_j^\pm)^\ast m_{N_i}  m_{N_j} P_{D_i} P_{D_j}^\ast T_\pm(DD^\ast) +\sum_{i,j=1}^2 v_i^\pm (v_j^\pm)^\ast m_{N_i}  m_{N_j} P_{D_i} P_{C_j}^\ast T_\pm(DC^\ast) \,\nn \\&\quad+ (D \leftrightarrow C )\bigg]\, .
\end{align}
In the second line of \eqref{eq:Mtot}, the suffix $D_j(C_j)$ stand for direct(cross) channel with $j^{\rm th}$ neutrino exchange, and in the last line we have introduced the following notations
\begin{align}
\mathcal{N} =\frac{1}{2}G_F^4 |V_{ud}|^2 |V_{qb}|^2 f_{\pi}^2,\,\,\,\ v_i^+=V_{\ell_1 N_i}V_{\ell_2 N_i},\,\,\,\ v_i^- = (v_i^+)^\ast \, ,
\end{align}
where $V_{qb} = V_{ub}$ for $\Lambda_b\to p^+\pi^+\ell^-\ell^-$, $V_{qb} = V_{cb}$ for $\Lambda_b\to \Lambda_c^+\pi^+\ell^-\ell^-$, and $f_\pi$ is the pion decay constant.
In the last line of \eqref{eq:Mtot}, the spin summed and averaged matrix element mod squared splits in to universal functions $T_\pm(XY^\ast)$, where $X(Y)=D,C$, and the functions $P_{X_j}$ which are functions of the masses $m_{N_1}, m_{N_2}$ and decay widths $\Gamma_{N_1}, \Gamma_{N_2}$ of the exchanged neutrinos 
\begin{align}
P_{D_j} = \frac{1}{(p_N^2-m_{N_j}^2)+i\Gamma_{N_j} m_{N_j}},\,\, P_{C_j} = \frac{1}{(p_N^{\prime 2}-m_{N_j}^2)+i\Gamma_{N_j} m_{N_j}}\, .
\end{align} 
Using \eqref{eq:Mtot}, the total decay widths can be conveniently written as
\begin{align}\label{eq:Br1}
\Gamma_{\mathcal{B}_1} 
&=(2-\delta_{\ell_1\ell_2}) \sum_{i,j=1}^2 v_i^+ (v_j^+)^\ast \Big(\widehat{\Gamma}(DD^\ast)_{ij}+\widehat{\Gamma}(CC^\ast)_{ij}+ \widehat{\Gamma}_+(DC^\ast)_{ij} + \widehat{\Gamma}_+(D^\ast C)_{ij} \Big)\, ,\\
\label{eq:Br2}
\Gamma_{\overline{\mathcal{B}}_1}
&=(2-\delta_{\ell_1\ell_2}) \sum_{i,j=1}^2 v_i^- (v_j^-)^\ast \Big(\widehat{\Gamma}(DD^\ast)_{ij}+\widehat{\Gamma}(CC^\ast)_{ij}+ \widehat{\Gamma}_-(DC^\ast)_{ij} + \widehat{\Gamma}_-(D^\ast C)_{ij} \Big)\,,
\end{align}
where the quantities $\widehat{\Gamma}_\pm$ are 
\begin{align}\label{eq:GHatpm}
\widehat{\Gamma}_\pm(XY^\ast)_{ij}&=\frac{\mathcal{N}}{2\mB1 2!}\int m_{N_i} m_{N_j}P_{X_i}P_{Y_j}^\ast T_{\pm}(XY^\ast) d\Phi_4,\,\,\, X,Y=C, D\, .
\end{align}
The expressions of $T_\pm(XY^\ast)$ and the requisite kinematics to evaluate these expressions are given in appendix \ref{sec:amps} and \ref{sec:kinem}, respectively, and the four-body phase space $d\Phi_4$ is given in appendix \ref{sec:phasespace}. In equations \eqref{eq:Br1}-\eqref{eq:Br2}, using the relation $T_+(XX^\ast) = T_-(XX^\ast)$ (see appendix \ref{sec:amps}) we have defined 
\begin{align}
\widehat{\Gamma}(XX^\ast)_{ij} \equiv \widehat{\Gamma}_+(XX^\ast)_{ij} = \widehat{\Gamma}_-(XX^\ast)_{ij}\, ,\quad X = D, C\, .
\end{align}
To physically interpret the terms, $\widehat{\Gamma}(XX^\ast)_{ij}$ are the contributions of $N_i$ exchange in the $X$ channel and the conjugate of $N_j$ exchange in the $X^\ast$ channel. The interference terms $\widehat{\Gamma}(XY^\ast)_{ij}$ are the contributions of $N_i$ exchange in the $X$ channel and the conjugate of $N_j$ exchange the $Y$ channel. Numerically, $D-C$ channel interference contributions $\widehat{\Gamma}(XY^\ast)_{ij}$ for $X\neq Y$, are insignificant compared to $\widehat{\Gamma}(XX^\ast)_{ij}$ and are ignored in our calculations.
In addition of the decay rates, the quantities of interests are their sum and differences
\begin{align}
\label{eq:sum}
\Gamma_{\mathcal{B}_1} + \Gamma_{\overline{\mathcal{B}}_1}&= 2 (2-\delta_{\ell_1 \ell_2})\Bigg[|V_{\ell_1 N_1}|^2 |V_{\ell_2 N_1}|^2 \bigg( \widehat{\Gamma}(DD^\ast)_{11} + \widehat{\Gamma}(CC^\ast)_{11}\bigg) \, \nn\\ &+ |V_{\ell_1 N_2}|^2 |V_{\ell_2 N_2}|^2 \bigg( \widehat{\Gamma}(DD^\ast)_{22} +  \widehat{\Gamma}(CC^\ast)_{22}\bigg) \, \nn\\ & + 2\cos(\theta_{21}) |V_{\ell_1 N_1}| |V_{\ell_2 N_1}| |V_{\ell_1 N_2}| |V_{\ell_2 N_2}| \bigg( \re\widehat{\Gamma}(DD^\ast)_{12} + \re\widehat{\Gamma}(CC^\ast)_{12} \bigg) 
\Bigg]\,,\\
\label{eq:diff}
\Gamma_{\mathcal{B}_1} - \Gamma_{\overline{\mathcal{B}}_1}&= 4 (2-\delta_{\ell_1 \ell_2}) |V_{\ell_1 N_1}| |V_{\ell_2 N_1}| |V_{\ell_1 N_2}| |V_{\ell_2 N_2}|\bigg[\sin(\theta_{21})\bigg(\im \widehat{\Gamma}(DD^\ast)_{12} + \im \widehat{\Gamma}(CC^\ast)_{12}\bigg) \bigg]\, ,
\end{align}
where the CP-odd phase, based on the convention adopted in \eqref{eq:phase}, is
\begin{align}
\begin{split}
\theta_{ij}&= \arg(V_{\ell_1N_i})+\arg(V_{\ell_2N_i})-\arg(V_{\ell_1N_j})-\arg(V_{\ell_2N_j})\, ,\\
&=(\phi_{1i}+\phi_{2i}-\phi_{1j}-\phi_{2j}),\,\,\,\ i, j=1,2\, .
\end{split}
\end{align}
A CP-even phase $\Delta\xi = \xi_1-\xi_2$ essential for CP violation is also present in the interference of $N_1$ and $N_2$ contributions
\begin{align}
\re \widehat{\Gamma}(XX^\ast)_{12}= \frac{\mathcal{N}}{2\mB1 2!}\int m_{N_1} m_{N_2}|P_{X_1}||P_{X_2}|\cos(\Delta\xi) T(XX^\ast) d_4^{\rm PS},\,\,\, X=C, D\, ,\\
\im \widehat{\Gamma}(XX^\ast)_{12}= \frac{\mathcal{N}}{2\mB1 2!}\int m_{N_1} m_{N_2}|P_{X_1}||P_{X_2}|\sin(\Delta\xi) T(XX^\ast) d_4^{\rm PS},\,\,\, X=D, C\, ,
\end{align}
where $\xi_{1,2}$ are given as
\begin{align}
\tan \xi_1=\frac{m_{N_1}\Gamma_{N_1}}{k_N^2-m_{N_1}^2},\,\,\,\,\,\ \tan \xi_2=\frac{m_{N_2}\Gamma_{N_2}}{k_N^2-m_{N_2}^2}\, ,
\end{align}
and $k_N^2=p_N^2$ for $D$-channel and $k_N^2=(p^{\prime}_N)^2$ for $C$-channel.\\
%

%%%%%%%%%%%%%%%%%%%%%%%%%%%%%%%%%%%%%%%%%%%%%%%%%%%%%%%%%%%%%%%%%%%%%%%%
%%%%%%%%%%%%%%%%%%%%%%%%%%%%%%%%%%%%%%%%%%%%%%%%%%%%%%%%%%%%%%%%%%%%%%%%
\section{Results \label{sec:num}}
Following the formalism in the previous section, we turn to numerical analysis with specific decay modes. At the LHC, about 5\% of the total $b$-hadrons produced are $\Lambda_b$ baryons, and both at the LHCb and CMS the muon reconstruction efficiency is comparatively higher than the other two charged leptons. We therefore are interested in the modes $\Lambda_b\to \Lambda_c\pi\mu\mu$ and $\Lambda_b\to p\pi\mu\mu$ channels. Since $\ell_1=\ell_2=\mu$, the CP-odd phase is $\theta_{21} = 2(\phi_{\mu 2} - \phi_{\mu 1})$. For numerical analysis, form factor parameterizing the $\Lambda_b^0\to\Lambda_c^+$ and $\Lambda_b^0\to p^+$ hadronic matrix elements are taken from the lattice QCD calculations \cite{Detmold:2015aaa}, and we take the decay constant of pion $f_\pi= 130.2(0.8)$ MeV from \cite{Aoki:2019cca}.

We also need to know the total decay widths of the heavy neutrinos $\Gamma_{N_{1,2}}$  as a function of their masses. For Majorana neutrino mass between $m_\pi + m_\mu < m_{N} < (\mB1-\mB2-m_{\mu})$, both purely leptonic as well as semi-hadronic decays may be relevant. For $m_N<1$ GeV, the decays to leptonic modes as well to light pseudo-scalar and vector mesons have been calculated in \cite{Atre:2009rg}. For higher values of $m_N$, decays to semi-hadronic mods are increasingly difficult due to the limited knowledge of the resonances. An inclusive approach based on quark-hadron duality was adopted in Ref. \cite{Helo:2010cw,Gribanov:2001vv} to calculate the widths of the semi-hadronic channel. For this analysis, we leave the decay width as a phenomenological parameter that can be measured by experiments. Following the analysis of \cite{Mejia-Guisao:2017nzx} we take the neutrino the lifetimes  $\tau_N = \hbar/\Gamma_N = [10, 100, 1000]$ps for numerical illustration.

We are interested in the signal of leptonic CP asymmetry
\begin{equation}\label{eq:main}
\mathcal{A}_{\rm CP} = \frac{\Gamma_{\mathcal{B}_1} - \Gamma_{\overline{\mathcal{B}}_1}}{\Gamma_{\mathcal{B}_1} + \Gamma_{\overline{\mathcal{B}}_1}}\, .
\end{equation}
The reason why this asymmetry will be present in the decay can be understood as follows. There are two interfering amplitudes coming from the two intermediate neutrinos $N_1$ and $N_2$. The interfering amplitudes have CP-odd phase $\theta_{21}$ that changes sign for the conjugate process. A CP-even phase $\Delta\xi$ comes from an absorptive part that is generated due to the interference of the two neutrino contributions and does not change sign in the conjugate process. In general, $\theta_{21}$ can be anything but a maximal $\mathcal{A}_{\rm CP}$ can be obtained for $\theta_{21} = \pi/2$ as can be seen from equation \eqref{eq:diff}. To understand the behavior of $\mathcal{A}_{\rm CP}$ with the neutrino mass, we note that  $\im[\widehat{\Gamma}(DD^\ast)_{ij}] \propto \im[P_{D_i}P_{D_j}^\ast]$. For our choices of the neutrino lifetime $\tau_N$ and the kinematically allowed neutrino mass $m_N$, the approximation $\Gamma_{N_j}\ll m_{N_j}$ is always valid so that
\begin{eqnarray}\label{eq:diag}
|P_{D_j}|^2 = \frac{\pi}{m_{N_j}\Gamma_{N_j}}\delta(p_N^2-m_{N_j}^2)\, ,\quad |P_{C_j}|^2 = \frac{\pi}{m_{N_j}\Gamma_{N_j}}\delta(p^{\prime 2}_N-m_{N_j}^2)\, ,
\end{eqnarray}
which yields 
\begin{equation}\label{eq:G1122}
\frac{\widehat{\Gamma}({XX^\ast})_{ii}}{\widehat{\Gamma}({XX^\ast})_{jj}}=\frac{\Gamma_{N_j}}{\Gamma_{N_i}}\, .
\end{equation}
When the mass difference between the neutrinos is such that  $\Gamma_{N_j}\ll\Delta m_N$ then we can write
\begin{eqnarray}\label{eq:PP}
\im[P_{D_1}P_{D_2}^\ast]\bigg|_{\Gamma_{N_j}\ll\Delta m_N} &= &\mathcal{P}\Big(\frac{1}{p_N^2-m_{N_1}^2} \Big)\pi\delta(p_N^2-m_{N_2}^2)- \mathcal{P}\Big(\frac{1}{p_N^2-m_{N_2}^2} \Big)\pi\delta(p_N^2-m_{N_1}^2)\nn\\
&=& \frac{\pi}{m_{N_2}^2-m_{N_1}^2}\Big(\delta(p_N^2-m_{N_1}^2) +\delta(p_N^2-m_{N_2}^2) \Big)\, ,\nn\\
&=& \frac{1}{y} \frac{2\pi}{(m_{N_1} +m_{N_2})(\Gamma_{N_1}+\Gamma_{N_1})}\Big(\delta(p_N^2-m_{N_1}^2) +\delta(p_N^2-m_{N_2}^2) \Big)\, ,
\end{eqnarray}
where $y=\Delta m_N/\Gamma_N$ and $\Gamma_N = (\Gamma_{N_1}+\Gamma_{N_2})/2$. 
This yields
\begin{align}
\frac{\im\widehat{\Gamma}({XX^\ast})_{12}}{\widehat{\Gamma}({XX^\ast})_{jj}} & =  \frac{\im\widehat{\Gamma}({XX^\ast})_{12}\big|_{\Gamma_{N_j}\ll\Delta m_N}}{\widehat{\Gamma}({XX^\ast})_{jj}} \frac{\im\widehat{\Gamma}({XX^\ast})_{12}}{\im\widehat{\Gamma}({XX^\ast})_{12}\big|_{\Gamma_{N_j}\ll\Delta m_N}} \, ,\nn\\
& = \frac{1}{y} \frac{4\pi}{(m_{N_1} +m_{N_2})(\Gamma_{N_1}+\Gamma_{N_1})} \frac{m_{N_j} \Gamma_{N_j}}{\pi} \eta\, ,
\end{align}
where the suppression factor
\begin{equation}
\eta \equiv \frac{\im\widehat{\Gamma}({XX^\ast})_{12}}{\im\widehat{\Gamma}({XX^\ast})_{12}\big|_{\Gamma_{N_j}\ll\Delta m_N}} \,,
\end{equation}
accounts for the departure from the approximation $\Gamma_{N_j}\ll\Delta m_N$ in the term $\im\widehat{\Gamma}({XX^\ast})$. Assuming the neutrinos to be almost degenerate $m_{N_1} \sim m_{N_2} = m_N$ 
\begin{align}\label{eq:relation1}
\frac{\im\widehat{\Gamma}({XX^\ast})_{12}}{\widehat{\Gamma}({XX^\ast})_{jj}} & = \frac{2\Gamma_{N_j}}{(\Gamma_{N_1}+\Gamma_{N_1})}  \frac{\eta(y)}{y}\, .
\end{align}
We define another factor
\begin{eqnarray}\label{eq:relation2}
\delta_j(y)&=&\frac{\re\widehat{\Gamma}({XX^\ast})_{12}}{\widehat{\Gamma}({XX^\ast})_{jj}}\, ,
\end{eqnarray}
which measures the interference of the two neutrinos in the real part of $\widehat{\Gamma}(XX^\ast)$. 
Following \eqref{eq:G1122} we get
\begin{equation}
\frac{\delta_1}{\delta_2}=\frac{\Gamma_{N_1}}{\Gamma_{N_2}}\, .
\end{equation}
Since we are considering same sign di-muon in the final state, $\eta$ and $\delta_j$ are same for the $D$ and $C$-channels which follows from the fact that
\begin{align}\label{eq:decay relation}
&\bar{\Gamma}({DD^\ast})_{jj}=\bar{\Gamma}({CC^\ast})_{jj},\,\ \re\bar{\Gamma}({DD^\ast})_{12}=\re\bar{\Gamma}({CC^\ast})_{12},\,\nn\\&\im\bar{\Gamma}({DD^\ast})_{12}=\im\bar{\Gamma}({CC^\ast})_{12}\, .
\end{align} 

The CP asymmetry can now be written in a convenient form as
\begin{eqnarray}
\mathcal{A}_{CP}=\frac{4\sin\theta_{21}}{\frac{|V_{\ell N_1}||V_{\ell N_1}|}{|V_{\ell N_2}||V_{\ell N_2}|}\Big(1+\frac{\Gamma_{N_2}}{\Gamma_{N_1}}\Big)+\frac{|V_{\ell N_2|}|V_{\ell N_2}|}{|V_{\ell N_1}||V_{\ell N_1}|}\Big(1+\frac{\Gamma_{N_1}}{\Gamma_{N_2}}\Big)+4\delta(y)\cos\theta_{21}}\frac{\eta(y)}{y}\, ,
\end{eqnarray}
where we define 
\begin{equation}
\delta(y)=\frac{\delta_1(y)+\delta_2(y)}{2}\, .
\end{equation}
For nearly degenerate neutrinos, it is natural to assume $|V_{\mu N_1}| \sim |V_{\mu N_2}| = |V_{\mu N}|$. This further simplifies the expression of $\mathcal{A}_{CP}$
\begin{eqnarray}\label{eq:acp}
\mathcal{A}_{CP}=\frac{4\sin\theta_{21}}{\Big(1+\frac{\Gamma_{N_2}}{\Gamma_{N_1}}\Big)+\Big(1+\frac{\Gamma_{N_1}}{\Gamma_{N_2}}\Big)+4\delta(y)\cos\theta_{21}}\frac{\eta(y)}{y}\, .
\end{eqnarray}
\begin{figure}[h!]
\begin{center}
\includegraphics[scale=0.29]{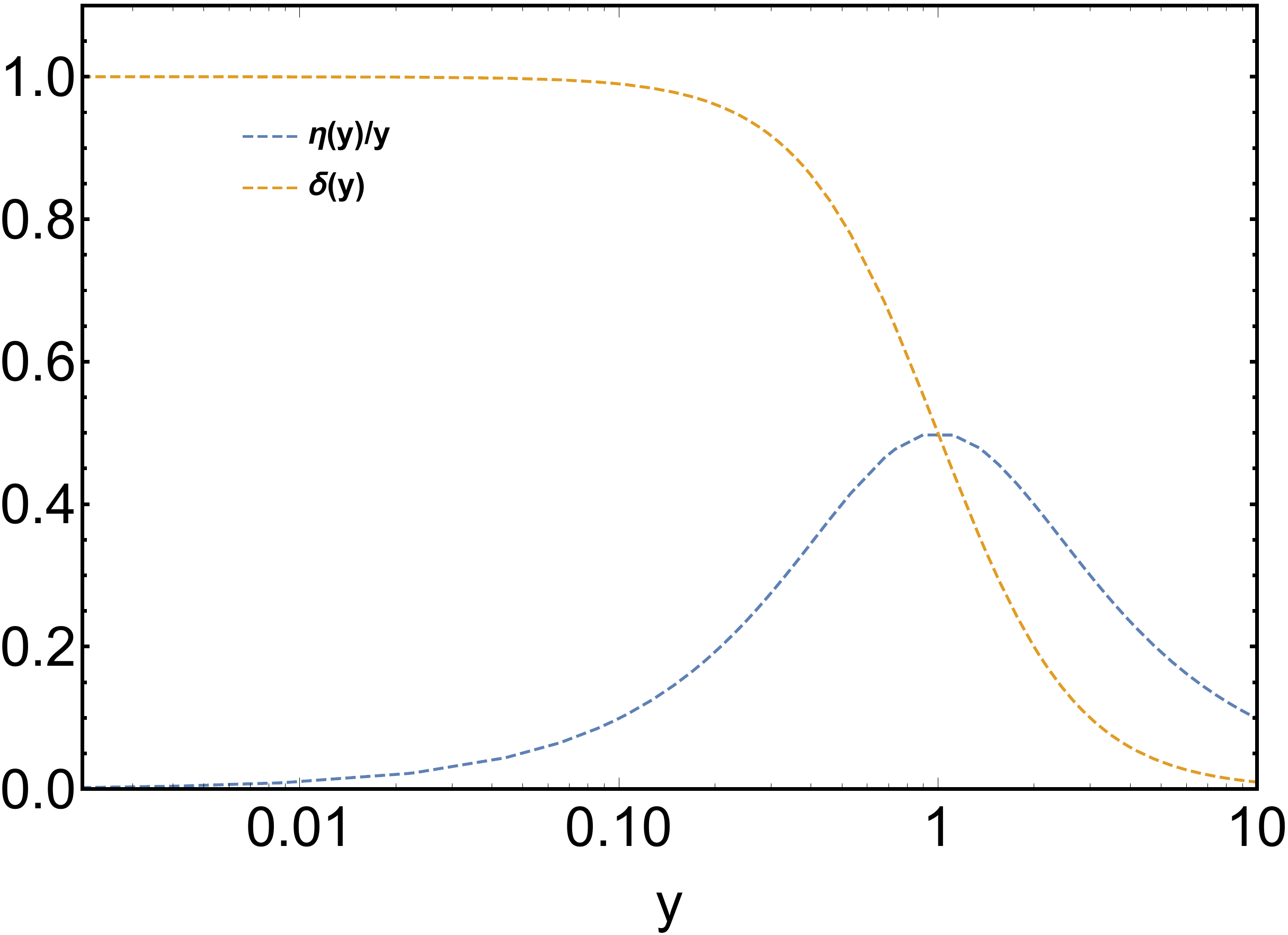}
\includegraphics[scale=0.3]{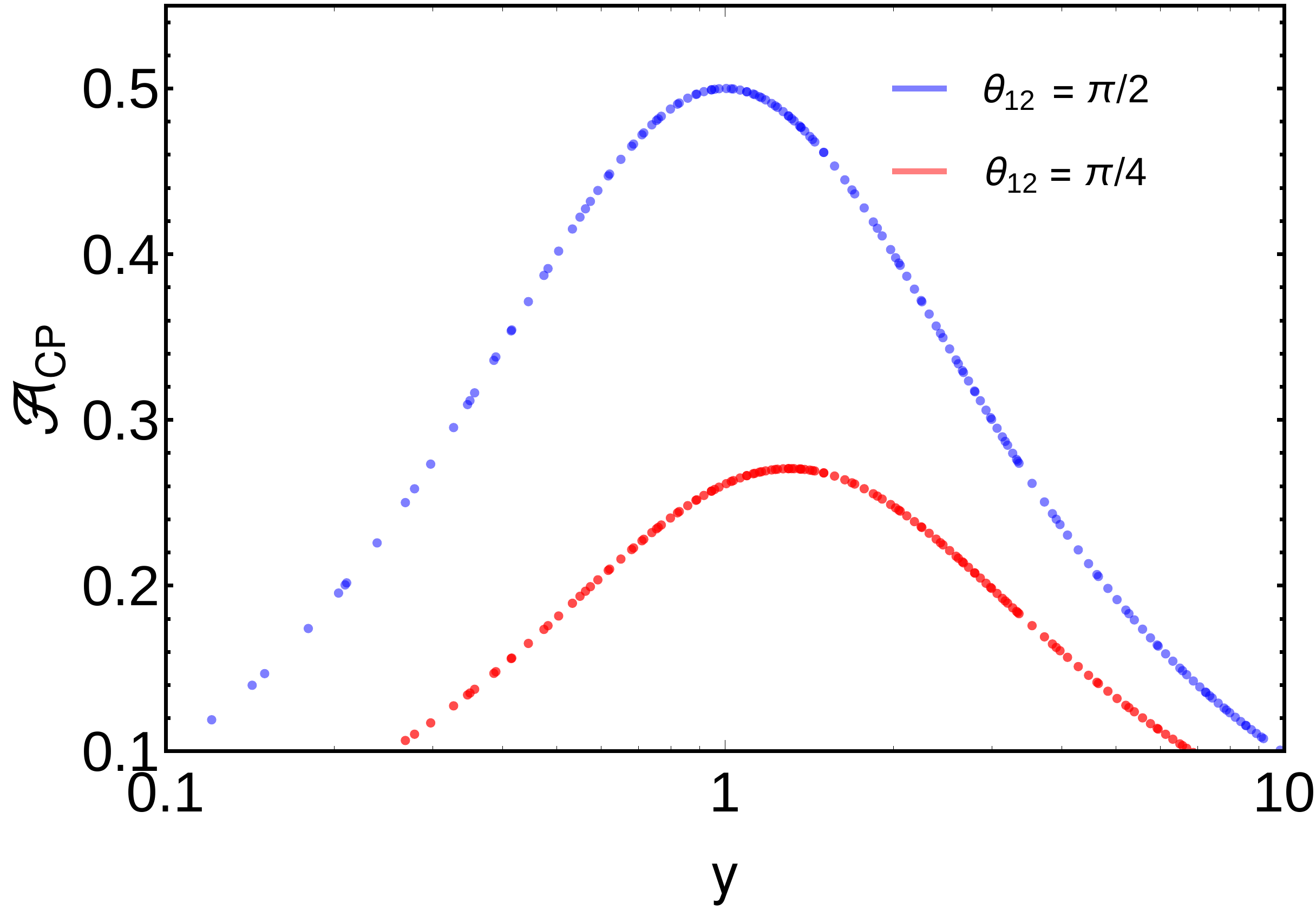}
\caption{The factors $\delta(y)$ and $\eta(y)/y$ as a function of $y=\Delta m_N/\Gamma_N$. The CP asymmetry observable $\mathcal{A}_{\rm CP}$ for $\Lambda_b\to\Lambda_c\pi\mu\mu$ is shown as a function of $y$ for different values of the weak phase. An identical plot is obtained for $\Lambda_b\to p\pi\mu\mu$. In these plots we have taken $|V_{\mu N_1}|^2 = |V_{\mu N_2}|^2 = 1  $\label{fig:delta}}
\end{center}
\end{figure}
\begin{figure}[h!]
	\begin{center}
		\includegraphics[scale=0.30]{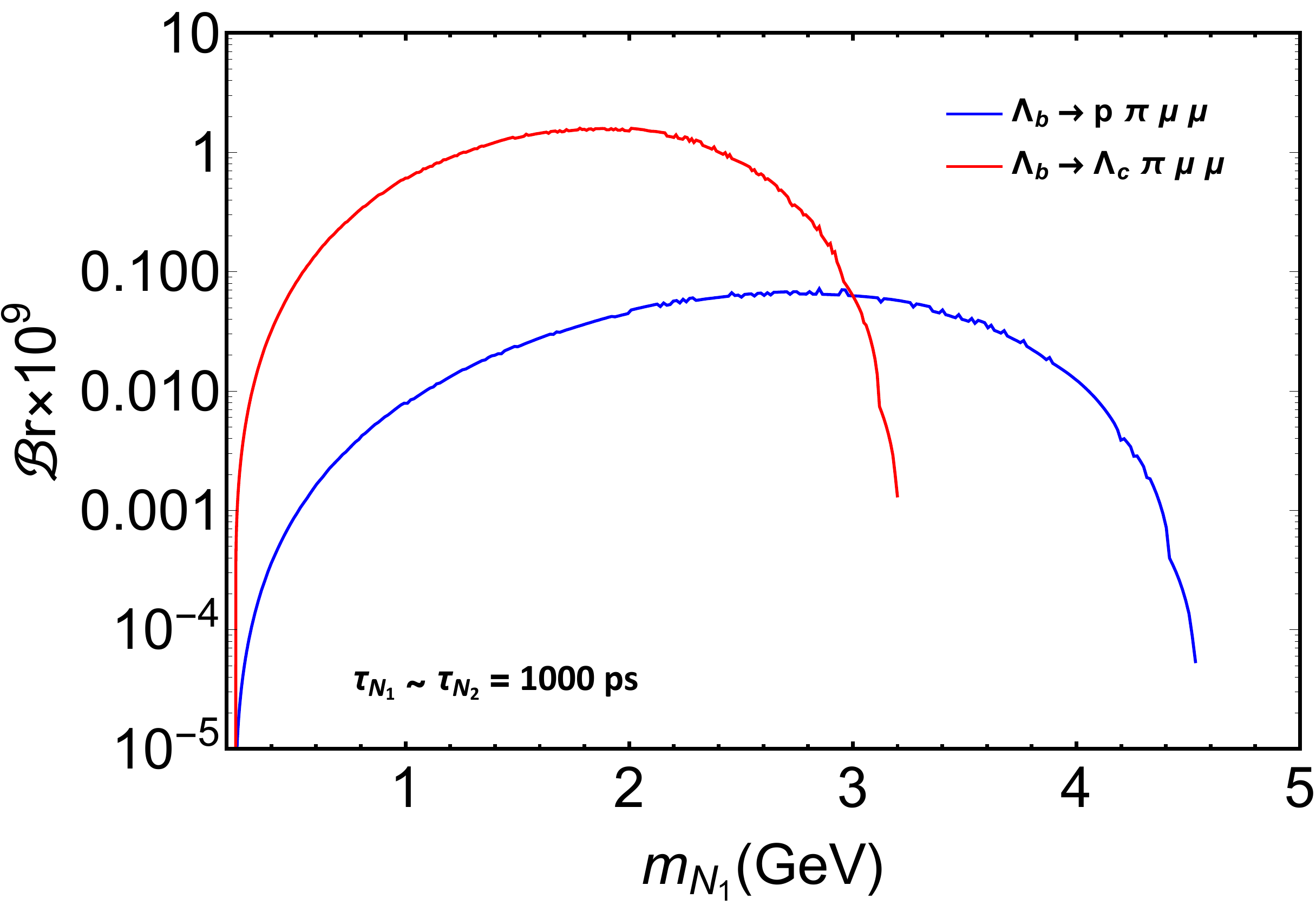}
		\includegraphics[scale=0.3]{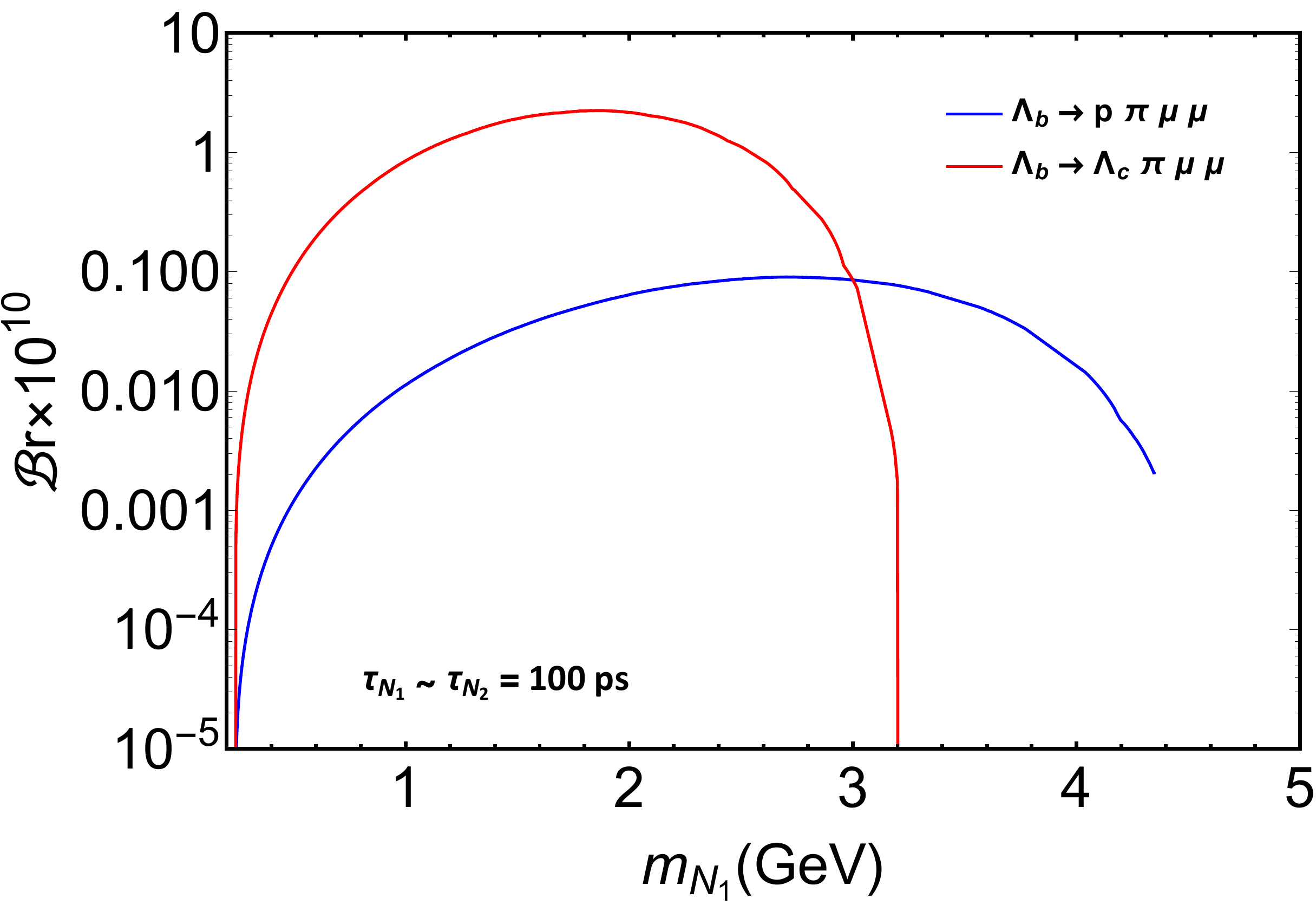}
		\caption{The branching ratios $\mathcal{B}(\Lambda_b\to\Lambda_c\pi\mu\mu)$ and $\mathcal{B}(\Lambda_b\to p\pi\mu\mu)$ for $|V_{\mu N_1}|^2 \sim |V_{\mu N_2}|^2 = |V_{\mu N}|^2 = 10^{-5} $, the weak phase $\theta_{21} = \pi/4$, the mass difference $\Delta m_N =10^{-15}$GeV, and the neutrino lifetimes $\tau_N=[100, 1000]$ps.\label{fig:Br}}
	\end{center}
\end{figure}

In figure \ref{fig:delta} we show the the suppression factor $\eta(y)/y$ and $\delta(y)$ as a function of $y$. This figure demonstrates that the $\mathcal{A}_{\rm CP}$ will be maximum for $y\sim1$, i.e., when $\Delta m_N \sim \Gamma_N$. In figure \ref{fig:delta} we also show the $\mathcal{A}_{\rm CP}$ for the $\Lambda_b\to \Lambda\pi\mu\mu$ mode for different values of $\theta_{21}$ and as a function of $y$. An identical plot is obtained for $\Lambda_b\to p\pi\mu\mu$. For a particular mode, the possibility to observe $\mathcal{A}_{\rm CP}$ does not depend entirely on its size but also depends on the decay rates. In figure \ref{fig:Br} we show the CP-averaged branching ratios 
\begin{eqnarray}
\mathcal{B}r(\mathcal{B}_1 \rightarrow \mathcal{B}_2\pi\mu \mu )=\frac{1}{2}\bigg(\mathcal{B}r(\mathcal{B}_1\to\mathcal{B}_2\pi^+\mu^-\mu^-) + \mathcal{B}r(\overline{\mathcal{B}}_1\to\overline{\mathcal{B}}_2\pi^-\mu^+\mu^+) \bigg)\, ,
\end{eqnarray}
of $\Lambda_b\to \Lambda_c\pi\mu\mu$ and $\Lambda_b\to p\pi\mu\mu$ as a function of sterile neutrino mass $m_{N_1}$ for neutrino lifetimes $\tau_{N_1}\sim \tau_{N_2}\sim 100$ps and 1000ps, $|V_{\mu N}|^2\sim 10^{-5}$, $\theta_{21}=\pi/4$, and the neutrino mass difference $\Delta m_N=10^{-15}$ GeV. We find that the branching ratio of $\Lambda_b\to \Lambda_c\pi\mu\mu$ can be within $10^{-10}-10^{-9}$ range, whereas $\mathcal{B}(\Lambda_b\to p\pi\mu\mu)\sim 10^{-12}-10^{-11}$ is suppressed due to small CKM element $V_{ub}$. These rates are within the reach of the future LHC sensitivity. For a detailed discussions of the number of events expected at the LHCb and CMS please see Ref. \cite{Mejia-Guisao:2017nzx}.

\begin{figure}[h!]
	\begin{center}
		\includegraphics[scale=0.30]{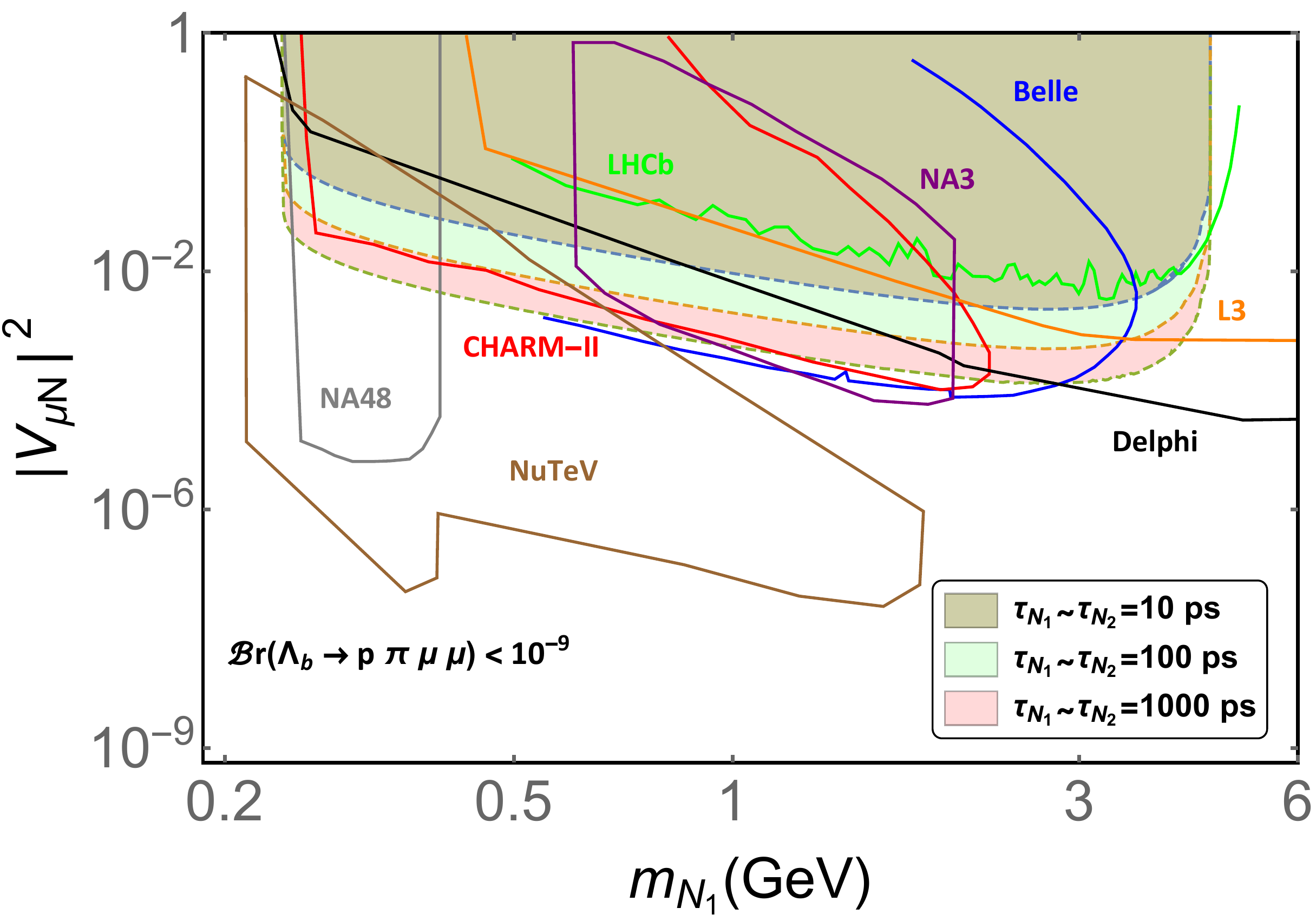}
		\includegraphics[scale=0.24]{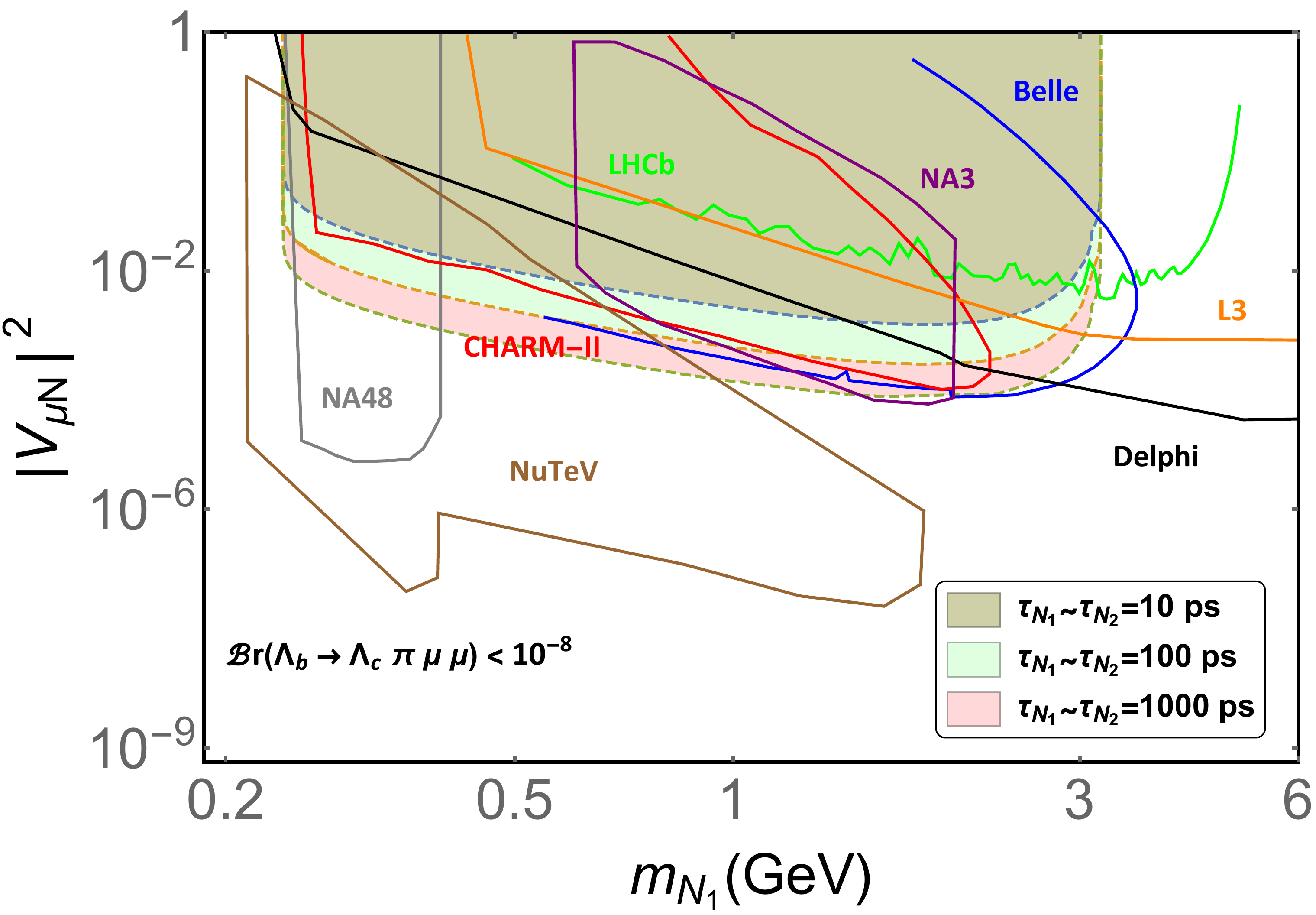}
		\includegraphics[scale=0.30]{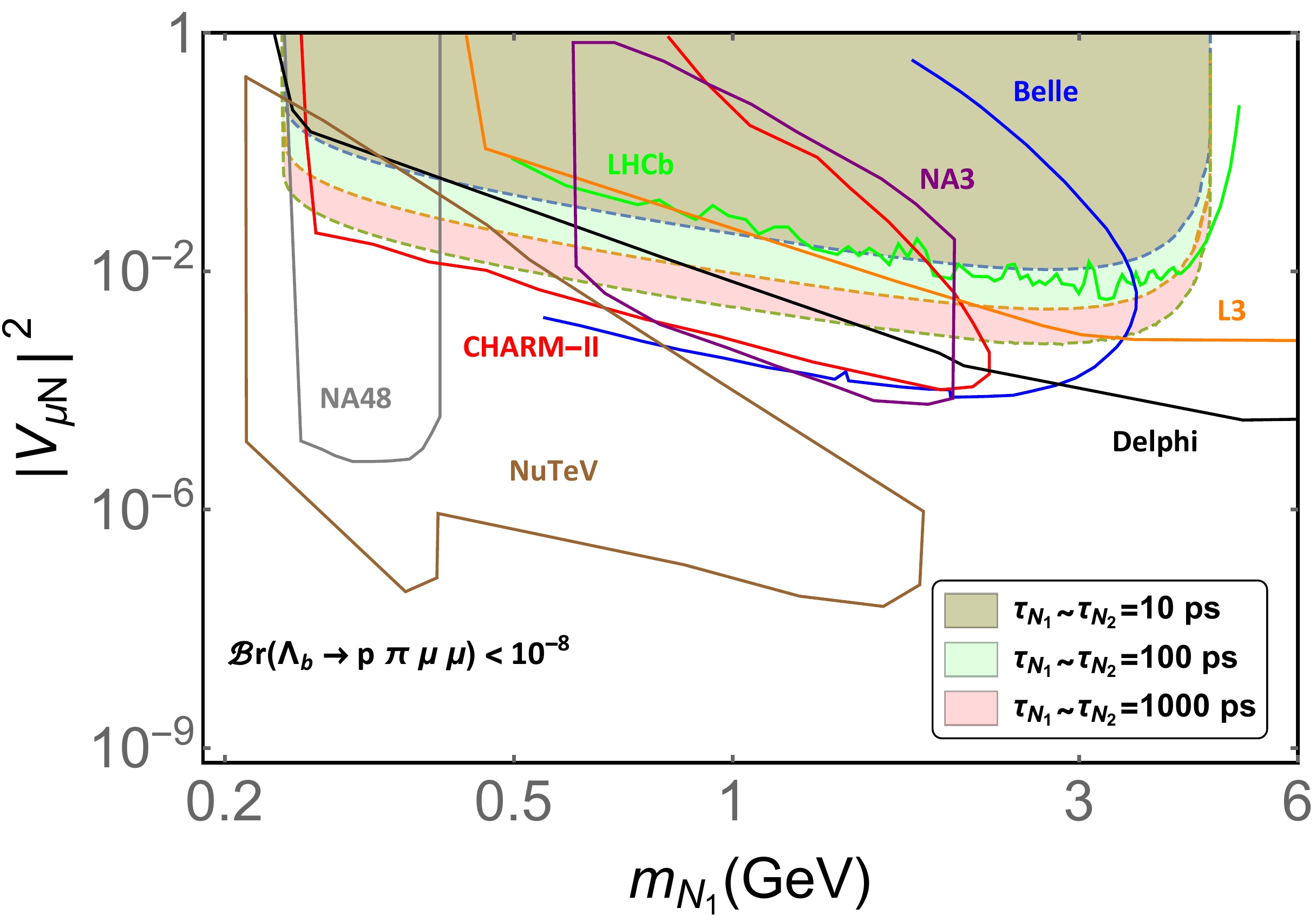}
		\includegraphics[scale=0.24]{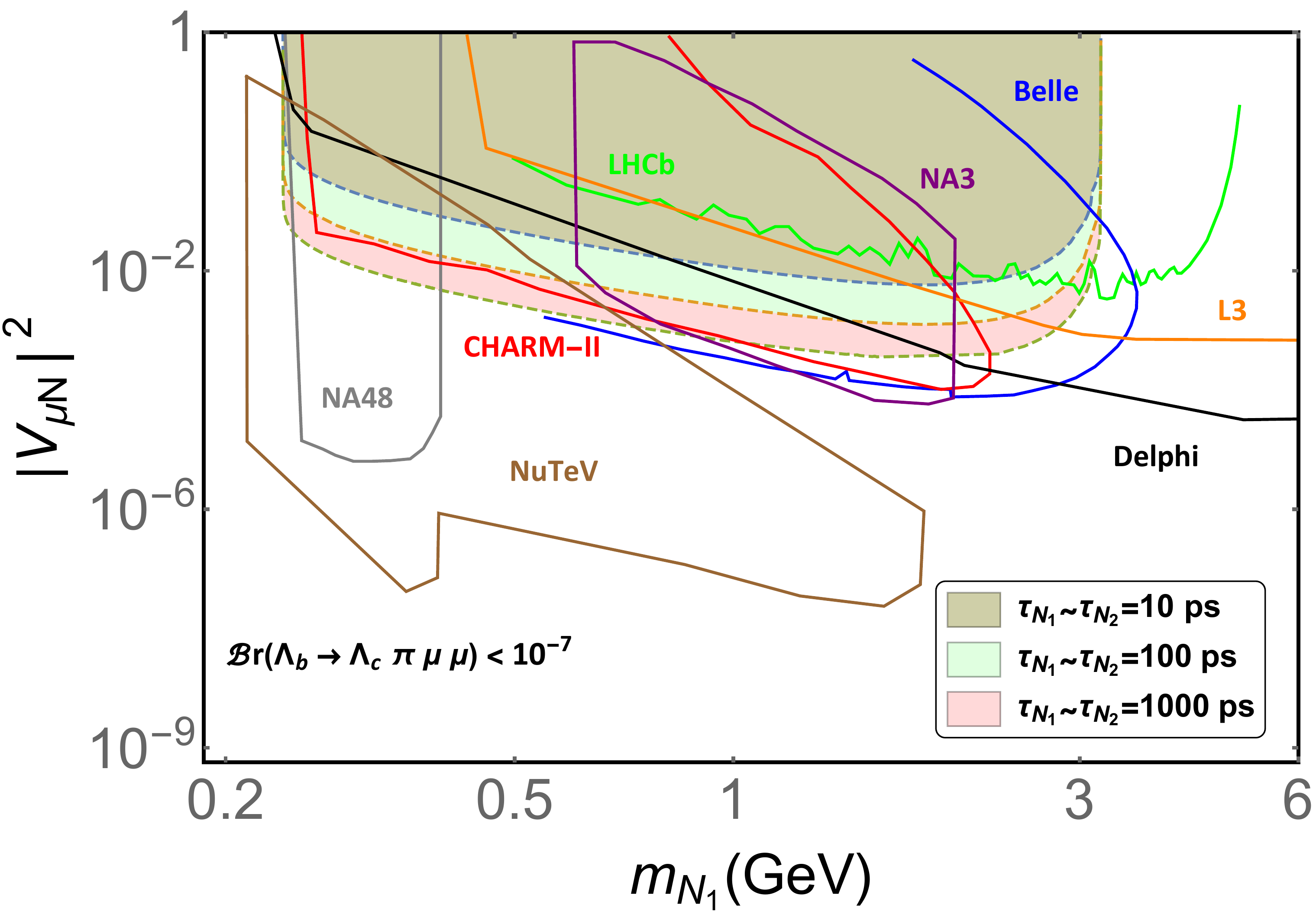}
		\caption{Exclusion regions on the $(m_{N_1}, |V_{\mu N}|^2)$ parameter space for $\mathcal{B}r(\Lambda_b\to p\pi\mu\mu)<10^{-8}, 10^{-9}$ and $\mathcal{B}r(\Lambda_b\to \Lambda_c\pi\mu\mu)<10^{-7},10^{-8}$ for different values of $\tau_N$, $\theta_{21} = \pi/4$, and $\Delta m_N = 10^{-15}$ GeV. \label{fig:const}}
	\end{center}
\end{figure}
Even if the decays are not fully observed, upper limits can be translated to bounds on the $m_N$ vs $|V_{\mu N}|^2$ parameter space.  In figures \ref{fig:const} we show the exclusion region in the ($m_N$,$|V_{\mu N}|^2$) plane obtained by assuming upper bounds $\mathcal{B}r(\Lambda_b\to p\pi\mu\mu)<10^{-8}, 10^{-9}$ and $\mathcal{B}r(\Lambda_b\to \Lambda_c\pi\mu\mu)<10^{-7},10^{-8}$ for different choices of the heavy neutrino lifetimes. The regions are shown in brown, light-green, and light-red correspond to exclusion regions obtained for $\tau_N=10{\rm ps}, 100{\rm ps}$ and $1000$ps respectively. To compare our bounds, in the figure \ref{fig:const} we also show the exclusion limits from LHCb \cite{Shuve:2016muy, Aaij:2014aba}, Belle \cite{Liventsev:2013zz}, L3 \cite{Adriani:1992pq}, Delphi \cite{Abreu:1996pa}, NA3 \cite{Badier:1985wg}, CHARM \cite{Vilain:1994vg}, NuTeV \cite{Vaitaitis:1999wq} and NA48 \cite{CERNNA48/2:2016tdo} experiments. These comparisons show that the LNV modes $\Lambda_b\to p\pi\mu\mu$ and $\Lambda_b\to p\pi\mu\mu$ can give complementary bounds on the sterile neutrino parameters. And with the possibility to observe CP asymmetry, the modes should be searched at the LHC.

%%%%%%%%%%%%%%%%%%%%%%%%%%%%%%%%%%%%%%%%%%%%%%%%%%%%%%%%%%%%%%%%%%%%%%%%
%%%%%%%%%%%%%%%%%%%%%%%%%%%%%%%%%%%%%%%%%%%%%%%%%%%%%%%%%%%%%%%%%%%%%%%%
\section{Summary \label{sec:summary}}
In this paper, we have studied lepton number violating baryonic decays $\Lambda_b\to\Lambda_c\pi\mu\mu$ and $\Lambda_b\to p \pi\mu\mu$ that are mediated by on-shell sterile Majorana neutrinos. The decays are studied in a model where there are two Majorana neutrinos. An interesting consequence of considering two Majorana neutrinos is that it gives rise to the possibility of CP violation in these modes. We find that appreciable CP asymmetry can be achieved if neutrinos are quasi-degenerate and the mass difference is of the order of decay widths. We have shown that in the absence of observation, upper limits to the branching ratios can give limits on the $m_N$ vs $|V_{\mu N}|^2$ parameter space that is comparable to limits obtained by other methods.

%%%%%%%%%%%%%%%%%%%%%%%%%%%%%%%%%%%%%%%%%%%%%%%%%%%%%%%%%%%%%%%
%%%%%%%%%%%%%%%%%%%%%%%%%%%%%%%%%%%%%%%%%%%%%%%%%%%%%%%%%%%%%%%
\section*{Acknowledgements}
The authors would like to thank Debajyoti Choudhury for fruitful discussions. DD acknowledges the DST, Govt. of India for the INSPIRE Faculty Fellowship (grant number IFA16-PH170). JD acknowledges the Council of Scientific and Industrial Research (CSIR), Govt. of India for JRF fellowship grant with File No. 09/045(1511)/2017-EMR-I. 
%%%%%%%%%%%%%%%%%%%%%%%%%%%%%%%%%%%%%%%%%%%%%%%%%%%%%%
%%%%%%%%%%%%%%%%%%%%%%%%%%%%%%%%%%%%%%%%%%%%%%%%%%%%%%

\appendix

%%%%%%%%%%%%%%%%%%%%%%%%%%%%%%%%%%%%%%%%%%%%%%%%%%%%%%%%%%%%%%%%%%%%%%
%%%%%%%%%%%%%%%%%%%%%%%%%%%%%%%%%%%%%%%%%%%%%%%%%%%%%%%%%%%%%%%%%%%%%%
\section{$\mathcal{B}_1\to \mathcal{B}_2^\mp\pi^\mp\ell_1^\pm\ell_2^\pm$ amplitudes \label{sec:amps}}
The effective Hamiltonian for the $\mathcal{B}_1\to \mathcal{B}_2\ell N$ decay and the subsequent decay of the intermediate neutrino $N\to \ell\pi$ are 
\begin{align}
& \mathcal{H}_{\rm eff}^{b\to q\ell N} = \frac{G_F}{\sqrt{2}} V_{qb}\bar{q}\gamma_\mu (1-\gamma_5)b \Bigg( \sum_{\ell=e}^\tau \sum_{i=1}^3 U_{\ell i} \bar{\nu}_i \gamma^\mu (1-\gamma_5)\ell + \sum_{\ell=e}^\tau \sum_{j=1}^{n} V_{\ell N_j} \bar{N_j^c} \gamma^\mu(1-\gamma_5)\ell \Bigg)+ {\rm h.c} \,, \\
& \mathcal{H}_{\rm eff}^{N_j\to\ell\pi} =\frac{G_F}{\sqrt{2}}V_{ud}\bar{d}\gamma_\mu(1-\gamma_5)u  \sum_{\ell=e}^\tau \sum_{j=1}^{n} V_{\ell N_j} \bar{\ell} \gamma^\mu(1-\gamma_5)N_j^c + {\rm h.c} \,.
\end{align}
The $\mathcal{B}_1\to\mathcal{B}_2\pi\ell\ell$ amplitudes for the $D$ and $C$ channel diagrams are 
\begin{align}
\mathcal{M}_{D_j}^\pm &= (G_F^2 V_{qb} M_{N_j}) \big( V_{\ell_1N_j}V_{\ell_2N_j} \big) P_{D_j} H_\nu ^\pm L^{\nu\alpha \pm}_{D}  J_\alpha^\pm \,,\\
\mathcal{M}_{C_j}^\pm &= (G_F^2 V_{qb} M_{N_j}) \big( V_{\ell_1N_j}V_{\ell_2N_j} \big) P_{C_j} H_\nu^\pm L^{\alpha\nu\pm}_{C}  J_\alpha^\pm \, ,
\end{align}
were the `+' correspond to the modes $\Lambda_b^0 \to (\Lambda_c^+,p^+)\pi^+ \ell_1^- \ell_2^-$ and the `-' correspond the conjugate modes. According to our convention, $H_\nu^+=H_\nu$ and $H_\nu^-=H_\nu^\ast$. The CKM elements $V_{qb} = V_{ub}$ for $\Lambda_b^0\to p^+\pi\ell\ell$ and $V_{qb} = V_{cb}$ for $\Lambda_b^0\to \Lambda_c^+\pi\ell\ell$ modes. The leptonic part of the amplitudes are
\begin{align}
& L^{\nu\alpha\pm}_D = \bar{u}_{\ell_1}(p_1) \gamma^\nu \gamma^{\alpha}(1\pm\gamma_5) v_{\ell_2}(p_2)\, ,\quad L^{\alpha\nu\pm}_C = \bar{u}_{\ell_1}(p_1) \gamma^{\alpha} \gamma^\nu (1\pm\gamma_5) v_{\ell_2}(p_2)\, .
\end{align}
The hadronic amplitudes $H^\mu$ are calculated using the form factor parametrization of $\mathcal{B}_1\to\mathcal{B}_2$ transitions from Ref.~\cite{Detmold:2015aaa}
\begin{align}\label{eq:VAhme1}
\langle \mathB2(k,s_k)|\bar{s}\gamma^\mu b |\mathB1(p,s_p)\rangle =& \bar{u}(k,s_k)\Bigg[f^V_t(q^2)(m_{\mathB1}-m_{\mathB2})\frac{q^\mu}{q^2}\nn\\ +& f^V_0(q^2) \frac{m_{\mathB1}+m_{\mathB2}}{s_+} \{p^\mu + k^\mu  - \frac{q^\mu}{q^2}(m_{\mathB1}^2 - m_{\mathB2}^2) \} \nn\\ + & f^V_\perp(q^2) \{ \gamma^\mu - \frac{2m_{\mathB2}}{s_+}p^\mu - \frac{2m_{\mathB1}}{s_+}k^\mu \} \Bigg]u(p,s_p)\, ,\\
%\end{align}
%
%
%\begin{align}
\label{eq:VAhme2}
\langle \mathB2(k,s_k)|\bar{s}\gamma^\mu\gamma_5 b |\mathB1(p,s_p)\rangle =& - \bar{u}(k,s_k) \gamma_5 \Bigg[ f_t^A(q^2) (m_{\mathB1} + m_{\mathB2}) \frac{q^\mu}{q^2} \nn\\ +& f_0^A(q^2) \frac{m_{\mathB1} - m_{\mathB2}}{s_-} \{p^\mu + k^\mu - \frac{q^\mu}{q^2} (m_{\mathB1}^2 - m_{\mathB2}^2) \} \nn\\ + & f_\perp^A(q^2) \{\gamma^\mu + \frac{2m_{\mathB2}}{s_-}p^\mu - \frac{2m_{\mathB1}}{s_-}k^\mu \}  \Bigg] u(p,s_p)\, .
\end{align}
Using \eqref{eq:VAhme1} and \eqref{eq:VAhme2} we can write the expression of $H^\mu$ as
\begin{align}
& H^\mu = \langle \mathB2(k,s_k)|\bar{c}\gamma^\mu(1-\gamma_5) b |\mathB1(p,s_p)\rangle \nn\\
&= \bar{u}(k,s_k)\Big(A_1 q^\mu+A_2 k^\mu+A_3\gamma^\mu+\gamma_5 \big\{A_4q^\mu+A_5 k^\mu+A_6\gamma^\mu\big\} \Big)u(p,s_p)\, ,
\end{align}
where the $q^2$ dependent functions $A_i$ can be written in terms of the form factors as
\begin{align}
& A_1 = f_t^V \frac{\mB1-\mB2}{q^2}+ f_0^V \frac{\mB1+\mB2}{s_+}  \bigg(1- \frac{\mmB1-\mmB2}{q^2} \bigg) - f_\perp^V \frac{2\mB2}{s_+} \, ,\\
& A_2 = 2 f_0^V  \frac{\mB1+\mB2}{s_+} - f_\perp^V \bigg(\frac{2\mB2}{s_+}+\frac{2\mB1}{s_+}\bigg) \, ,\\
& A_3=f_\perp^V\, ,\\
& A_4 = f_t^A \frac{\mB1+\mB2}{q^2} + f_0^A \frac{\mB1-\mB2}{s_-}  \bigg(1- \frac{\mmB1-\mmB2}{q^2} \bigg) + f_\perp^A \frac{2\mB2}{s_-} \, ,\\
& A_5 = 2 f_0^A  \frac{\mB1-\mB2}{s_-} + f_\perp^A \bigg(\frac{2\mB2}{s_-} - \frac{2\mB1}{s_-}\bigg) \, ,\\
& A_6=f_\perp^A\, .
\end{align}
Finally, the amplitudes for the pion production is 
\begin{align}
&J_{\alpha}^+ =\braket{\pi^+(k)|\bar{u}\gamma_\mu(1-\gamma_5)d|0}=i k_\mu V_{ud} f_{\pi}\, ,\quad 
J_{\alpha}^- =-ik_\mu V_{ud}^\ast f_{\pi}=(J_{\alpha}^+)^{\dagger}\, .
\end{align}

The matrix element mod-squared after summing over the final spins and averaging over the initial spin is given in equation \eqref{eq:Mtot}. The quadratic terms $T_\pm(XY^\ast)$ and $T(XY^\ast)$ given on \eqref{eq:Mtot} are
\begin{align}\label{eq:Tplmi}
T_\pm(DD^\ast) & = \sum_{spins}[H_\nu^\pm(H_\rho^\pm)^\ast] \sum_{spins}[L_D^{\nu\alpha\pm}(L_D^{\rho\beta\pm})^\ast] k_{\pi\alpha} k_{\pi\beta}\\
T_\pm(CC^\ast) & =\sum_{spins}[H_\nu^\pm(H_\rho^\pm)^\ast] \sum_{spins}[L_C^{\alpha\nu\pm}(L_C^{\beta\rho\pm})^\ast] k_{\pi\alpha} k_{\pi\beta}\\
T_\pm(DC^\ast) &= \sum_{spins}[H_\nu^\pm(H_\rho^\pm)^\ast] \sum_{spins}[L_D^{\nu\alpha\pm}(L_C^{\beta\rho\pm})^\ast] k_{\pi\alpha} k_{\pi\beta}\\
T_\pm(D^\ast C) &= \sum_{spins}[(H_\nu^\pm)^\ast H_\rho^\pm] \sum_{spins}[(L_D^{\nu\alpha\pm})^\ast L_C^{\beta\rho\pm}] k_{\pi\alpha} k_{\pi\beta}\\
T(DD^\ast) &\equiv T_+(DD^\ast) =T_-(DD^\ast),\,\,\, T(CC^\ast) \equiv T_+(CC^\ast) =T_-(CC^\ast)\,, \\
T_+(DC^\ast) &=T_-(D^\ast C) ,\,\,\,T_-(DC^\ast) =T_+(D^\ast C). 
\end{align}
In the appendix \ref{sec:kinem} we describe the kinematics required to calculate the momentum dot products required to evaluate the quadratic terms $T_\pm(XY^\ast)$.

%%%%%%%%%%%%%%%%%%%%%%%%%%%%%%%%%%%%%%%%%%%%%%%%%%%%%%%%%%%%%%%
%%%%%%%%%%%%%%%%%%%%%%%%%%%%%%%%%%%%%%%%%%%%%%%%%%%%%%%%%%%%%%%
\section{Kinematics \label{sec:kinem}}
As mentioned in the texts, the contribution coming from the interference of the direct and cross channel diagrams are negligibly small and neglected in our calculations. Therefore, kinematics of the direct and cross channel can be evaluated independently. In this section, we work out the kinematics for the direct channel. The cross channel can be obtained trivially from the results presented here. 

Referring to diagrams \ref{fig:Feyn}, in this section we work out the kinematics of $\mathcal{B}^0_1(\pB1)\to \mathcal{B}_2(\pB2)\pi^+(p_\pi)\ell_1^-(p_1)\ell_2^-(p_2)$ decay in the $\mathB1(p_{\mathB1})$ rest frame ($\mathB1$-RF). In this frame the four momentum of $\mathB2(p_{\mathB2})$ and $W_1(q)$ are
\begin{align}
& p^{\mathB1\text{-RF}}_{\mathB2} \equiv (m_{\mathB1}-E^{\mathB1 \text{-RF}}_q, 0, 0, {\bf p}^{\mathB1 \text{-RF}}_{\mathB2})\, ,\\
& q^{\mathB1 \text{-RF}} \equiv (E^{\mathB1 \text{-RF}}_q, 0, 0, -{\bf p}^{\mathB1 \text{-RF}}_{\mathB2})\, ,
\end{align}
where the $q^0$ and the ${\bf p}^{\mathB1 \text{-RF}}_{\mathB2}$ are 
\begin{align}
& E^{\mathB1 \text{-RF}}_q = \frac{\mmB1 + q^2 - \mmB2}{2\mB1}\, ,\quad |{\bf p}^{\mathB1 \text{-RF}}_{\mathB2}| = \frac{\sqrt{\lambda(\mmB1,\mmB2,q^2)}}{2\mB1}\, ,
\end{align}
and $\lambda(a,,b,c) = a^2 + b^2 + c^2 -2(ab+bc+ca)$. In the $W^-$-RF, we define $\theta_1$ as the angle made by $\ell_1$ with respect to the $\mathB2$. {\emph i.e.,} in the $+\hat{z}$ direction. The four momentum of $\ell_1(p_1)$ and $N(p_N)$ reads
\begin{align}
& p_1^{W^-\text{-RF}} = (E_1^{W^- \text{-RF}}, |{\bf p}_1^{W^-\text{-RF}}|\sin\theta_1, 0, |{\bf p}_1^{W^-\text{-RF}}|\cos\theta_1)\, ,\\
& p_N^{W^-\text{-RF}} = (\sqrt{q^2}-E_1^{W^-\text{-RF}}, -|{\bf p}_1^{W^-\text{-RF}}|\sin\theta_1, 0, -|{\bf p}_1^{W^-\text{-RF}}|\cos\theta_1)\, ,
\end{align}
where $E_1^{W^-\text{-RF}}$ and ${\bf p}_1^{W^-\text{-RF}}$ are given as 
\begin{align}
E_1^{W^-\text{-RF}} = \frac{q^2+m_1^2-p_N^2}{2\sqrt{q^2}}\, ,\quad |{\bf p}_1^{W^-\text{-RF}}| = \frac{\sqrt{\lambda(q^2, m_1^2,p_N^2)}}{2\sqrt{q^2}}\, .
\end{align} 
The Lorentz boost matrix to transform four vectors from the $W^-\text{-RF}$ to the $\mathB1\text{-RF}$ reads 
\begin{equation}
\Lambda_{W^- \to \mathB1} = \begin{pmatrix}
\gamma_1 & -\gamma_1\beta_{1x} & -\gamma_1\beta_{1y} & -\gamma_1\beta_{1z}\\ 
-\gamma_1\beta_{1x} & 1+(\gamma_1-1)\frac{\beta^2_{1x}}{\vec{\beta}_1^2} &  (\gamma_1-1)\frac{\beta_{1x}\beta_{1y}}{\vec{\beta}_1^2} & (\gamma_1-1)\frac{\beta_{1x}\beta_{1z}}{\vec{\beta}_1^2}\\
-\gamma_1\beta_{1y} & (\gamma_1-1)\frac{\beta_{1x}\beta_{1y}}{\vec{\beta}_1^2} &  1+(\gamma_1-1)\frac{\beta^2_{1y}}{\vec{\beta}_1^2} & (\gamma_1-1)\frac{\beta_{1y}\beta_{1z}}{\vec{\beta}_1^2}\\
-\gamma_1\beta_{1z} & (\gamma_1-1)\frac{\beta_{1x}\beta_{1z}}{\vec{\beta}_1^2} &  (\gamma_1-1)\frac{\beta_{1y}\beta_{1z}}{\vec{\beta}_1^2} & 1+(\gamma_1-1)\frac{\beta^2_{1z}}{\vec{\beta}_1^2}\\
\end{pmatrix}\, ,
\end{equation}
where the velocity $\vec{\beta}_1$ is the velocity of the $W^-(q)$ as seen in the $\mathB1\text{-RF}$ and
\begin{align}
& \gamma_1 = \frac{1}{\sqrt{1-\vec{\beta}_1^2}}\,,\quad \beta_{1x} = 0\, ,\quad \beta_{1y} = 0\, , \quad \beta_{1z} = \frac{|{\bf p}^{\mathB1\text{-RF}}_{\mathB2}|}{E^{\mathB1\text{-RF}}_q}\,.
\end{align}

In the heavy neutrino rest frame $N\text{-RF}$, the four momentum of $\ell_2(p_2)$ and the $W^+(p_\pi)$ are given as
\begin{align}
& p_2^{N{\rm-RF}} \equiv (E_2^{N{\rm-RF}}, |{\bf p}_2^{N{\rm-RF}}|\sin\theta_2\cos\phi,  |{\bf p}_2^{N{\rm-RF}}|\sin\theta_2\sin\phi,  |{\bf p}_2^{N{\rm-RF}}|\cos\theta_2 )\, ,\\
& p_\pi^{N{\rm-RF}} \equiv (E_\pi^{N{\rm-RF}}, -|{\bf p}_2^{N{\rm-RF}}|\sin\theta_2\cos\phi,  -|{\bf p}_2^{N{\rm-RF}}|\sin\theta_2\sin\phi,  -|{\bf p}_2^{N{\rm-RF}}|\cos\theta_2 )\, ,
\end{align}
where $\theta_2$ is the angle made the lepton $\ell_2$ with respect to the $+\hat{z}$ direction. The angle $\phi$ is made by the plane containing $\ell_2$ and $W^+$ with respect to the plane containing $\ell_1$ and $N$ as seen in the ${\mathB1{\rm -RF}}$ and is defined as
\begin{align}
& \cos\phi = (\hat{p}_1^{\mathB1{\rm -RF}} \times \hat{p}_N^{\mathB1{\rm -RF}}). (\hat{p}_2^{\mathB1{\rm -RF}} \times \hat{p}_{W^+}^{\mathB1{\rm -RF}})\, ,\\
& \sin\phi = -\bigg[(\hat{p}_1^{\mathB1{\rm -RF}} \times \hat{p}_N^{\mathB1{\rm -RF}})\times(\hat{p}_2^{\mathB1{\rm -RF}} \times \hat{p}_{W^+}^{\mathcal{B}_1{\rm -RF}})\bigg].\hat{p}_N^{\mathB1{\rm -RF}}\, .
\end{align}
The energies $E_2^{N{\rm-RF}}, E_\pi^{N{\rm-RF}}$ and three momentum $|{\bf p}_2^{N{\rm-RF}}|$ are
\begin{equation}
E_2^{N{\rm-RF}} = \frac{p_N^2+m_2^2 - p_\pi^2}{2\sqrt{p_N^2}}\, ,\quad E_\pi^{N{\rm-RF}} = \sqrt{p_N^2}-E_2^{N{\rm-RF}},\quad |{\bf p}_2^{N{\rm-RF}}| = \frac{\sqrt{\lambda(p_N^2, m_2^2, p_\pi^2)}}{2\sqrt{p_N^2}}\, .
\end{equation}
The Lorentz boost required to go from the heavy neutrino-rest frame to the $W^-$ rest-frame is given as
\begin{equation}
\Lambda_{N \to W^-} = \begin{pmatrix}
\gamma_2 & -\gamma_2\beta_{2x} & -\gamma_2\beta_{2y} & -\gamma_2\beta_{2z}\\ 
-\gamma_2\beta_{2x} & 1+(\gamma_2-1)\frac{\beta^2_{2x}}{\vec{\beta}_2^2} &  (\gamma_2-1)\frac{\beta_{2x}\beta_{2y}}{\vec{\beta}_2^2} & (\gamma_2-1)\frac{\beta_{2x}\beta_{2z}}{\vec{\beta}_2^2}\\
-\gamma_2\beta_{2y} & (\gamma_2-1)\frac{\beta_{2x}\beta_{2y}}{\vec{\beta}_2^2} &  1+(\gamma_2-1)\frac{\beta^2_{2y}}{\vec{\beta}_2^2} & (\gamma_2-1)\frac{\beta_{2y}\beta_{2z}}{\vec{\beta}_2^2}\\
-\gamma_2\beta_{2z} & (\gamma_2-1)\frac{\beta_{2x}\beta_{2z}}{\vec{\beta}_2^2} &  (\gamma_2-1)\frac{\beta_{2y}\beta_{2z}}{\vec{\beta}_2^2} & 1+(\gamma_2-1)\frac{\beta^2_{2z}}{\vec{\beta}_2^2}\\
\end{pmatrix}\, ,
\end{equation}
where the velocity $\vec{\beta}_2$ is the velocity of the $N$ as seen in the $W^-$-RF
\begin{align}
& \gamma_2 = \frac{1}{\sqrt{1-\vec{\beta}_2^2}}\,,\quad \beta_{2x} =  \frac{|{\bf p}^{W^-\rm-RF}_{N}|}{E^{W^-\rm-RF}_N}\sin\theta_1\, ,\quad \beta_{2y} = 0\, , \quad \beta_{2z} =  \frac{|{\bf p}^{W^-\rm-RF}_{N}|}{E^{W^-\rm-RF}_N}\cos\theta_1\, .
\end{align}
 
%%%%%%%%%%%%%%%%%%%%%%%%%%%%%%%%%%%%%%%%%%%%%%%%%%%%%%%%%%%%%%%
%%%%%%%%%%%%%%%%%%%%%%%%%%%%%%%%%%%%%%%%%%%%%%%%%%%%%%%%%%%%%%%
\section{Phase space \label{sec:phasespace}}
The differential width for the four-body final state is
\begin{equation}
d\Gamma = \frac{1}{2\mB1} |\mathcal{M}|^2 d\Phi_4\bigg(\mathB1\to\mathB2\ell_1\ell_2\pi\bigg)\, ,
\end{equation}
The four-body phase-space can be split up as
\begin{align}
d\Phi_4(\mathB1\to\mathB2\ell_1\ell_2\pi) &= d\Phi_3(\mathB1\to\mathB2\ell_1 N) \frac{dp_N^2}{2\pi} d\Phi_2(N\to \ell_2\pi)\, ,\quad\text{for $D$-channel}\,,\\
&= d\Phi_3(\mathB1\to\mathB2\ell_2 N) \frac{dp_N^2}{2\pi} d\Phi_2(N\to \ell_1\pi)\, ,\quad\text{for $C$-channel}\,,
\end{align}
The expressions of different phase space are given below
\begin{align}
& d\Phi_3(\mathB1\to\mathB2\ell_1 N) = \frac{\sqrt{\lambda(1,\mmB2/\mmB1,q^2/\mmB1)}}{4\pi}\frac{\sqrt{\lambda(1,m_{\ell_1}^2/q^2,p_N^2/q^2)}}{(8\pi)^2} \int dq^2 d\cos\theta_1\, ,\\
& d\Phi_2(N\to \ell_2\pi) = \frac{\sqrt{\lambda(1,m_{\ell_2}^2/p_N^2,p_\pi^2/p_N^2)}}{8\pi} \int \frac{d\cos\theta_2}{2} \frac{d\phi}{2\pi}\, ,
\end{align}

The limits of the integration of the angles are as follows 
\begin{align}
-1\leq \cos\theta_1\leq 1\, ,\quad -1\leq \cos\theta_2\leq 1\, ,\quad 0\leq \phi\leq 2\pi\, .\quad
\end{align}

%%%%%%%%%%%%%%%%%%%%%%%%%%%%%%%%%%%%%%%%%%%%%%%%%%%%%%%%%%%%%%%%%%%
%%%%%%%%%%%%%%%%%%%%%%%%%%%%%%%%%%%%%%%%%%%%%%%%%%%%%%%%%%%%%%%%%%%

\end{document}